# Intelligent Throughput-based Sleep Control Algorithm for the 5G Dense Heterogeneous Cellular Networks

Topside E. Mathonsi[1] and Tshimangadzo M. Tshilongamulenzhe[2]

[1,2]Department of Information Technology, Faculty of Information and Communication Technology, Tshwane University of Technology, Pretoria, South Africa
MathonsiTE@tut.ac.za and
TshilongamulenzheTM@tut.ac.za

**Abstract.** In the recent past, many mobile/telecom operators have seen a continuously growing demand for ubiquitous high-speed wireless access and an unprecedented increase in connected wireless devices. As a result, we have seen explosive growth in traffic volumes and a wide range of QoS requirements. The Fifth generation (5G) heterogeneous cellular networks (HetNets) have been developed by different mobile operators to achieve the growing mass data capacity and to reconnoiter the energy efficiency guaranteed trade-off between throughput QoS requirements and latency performance. However, existing energy efficiency algorithms do not satisfy the throughput QoS requirements such as reduced latency and packet loss, longer battery lifetime, reliability, and high data rates with regards to the three components of energy consumption of the 5G radio access network (RANs) that dominate the overall mobile communication networks. In addition, real-time traffic types such as voice and video require a high computational load at the terminal side which has an undesirable impact on energy/battery lifetime which further affects the throughput QoS performance such as reduced packet loss, longer battery lifetime, reliability, and high data rates. As a result, this paper proposed an Intelligent Throughput-based Sleep Control (ITSC) algorithm for throughput QoS and energy efficiency enhancement in 5G dense HetNets. In the proposed ITSC algorithm, a deep neural network (DNN) was used to determine the cell capacity ratio for the small base stations (SBSs). Hence, the SBSs cell capacity ratio was employed as decision criteria to put the SBSs into a sleep state. Furthermore, transferable payoff coalitional game theory was used in order to ensure real-time applications have a higher priority over non-real time applications. Numerous Network Simulator 2 (NS-2) results confirmed that the proposed ITSC algorithm reduced packet loss and produced better QoS. Moreover, the ITSC algorithm provided a longer battery lifetime, reliability, and high data rates for real-time traffic. The network throughput was improved as a result.

## 1 Introduction

In the recent past, wireless technologies have been growing actively all around the world and it has demonstrated exceptional changes in the way mobile and wireless communication systems are used. Hence, many mobile/ telecom operators have seen a continuously growing demand for ubiquitous high-speed wireless access and the rapid increase in connected wireless devices. As a result, we have seen a rapid growth in traffic volumes and a wide range of QoS requirements [1]-[35]. Continually, the Fifth generation (5G) heterogeneous cellular networks (HetNets) have been developed by different telecom operators to achieve the growing mass data capacity and to reconnoiter the energy efficiency guaranteed trade-off between throughput QoS requirements and latency performance [12]-[35].

It was predicted that by 2020, 5G HetNets will facilitate the connection of billions of wireless devices, i.e., wearable devices, sensors, drones, and cars for our smart homes, smart cities, smart and safer vehicles, telesurgery, advanced security, e-health, and smart schools (see Fig. 1). As a result, the 5G HetNets is providing a safer and more efficient place to live in the world as illustrated in Fig. 1 [1]-[35]. The connection of billions of wireless devices led to 5G dense HetNets as illustrated in Fig. 1. The 5G dense HetNets is made up of multiple radio base stations (RBSs)/small base stations (SBSs) to increase the coverage and system capacity which led to a high number of network elements. Hence, significantly increases the power consumption and lessening the energy efficiency in downlink cellular networks. As a result, balancing the trade-off between energy efficiency, throughput QoS requirements and latency performance has been a challenging factor for many telecom operators and has emerged as a new key pillar/protuberant figure of merit. This is due to the environmental concerns and operational costs for these telecom operators.

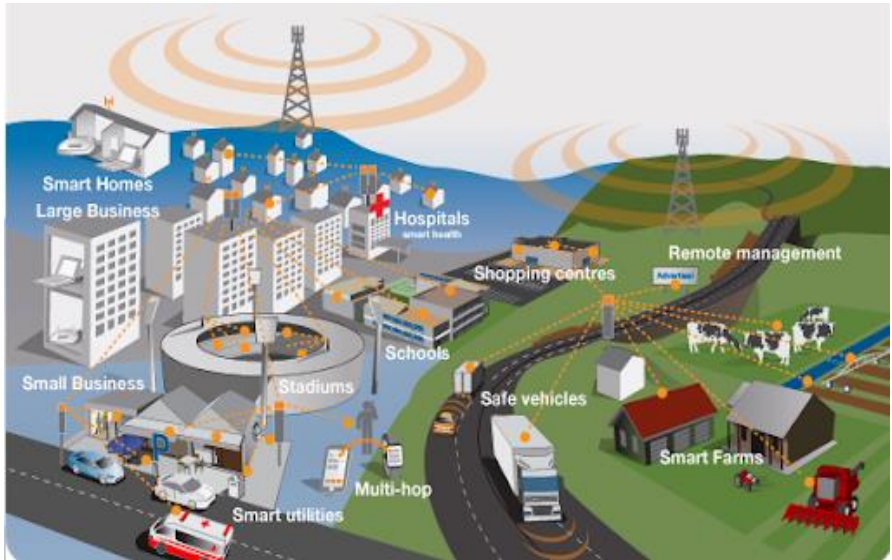

**Fig. 1.** A typical 5G dense HetNets [19]

Moreover, these three elements each contain two parts specifically, the dynamic power consumption part, which differs according to the network traffic load, and the fixed power consumption part, which is autonomous from the network traffic load. Continually, real-time traffic types such as voice and video require high computational load at the terminal side which have an undesirable impact on energy/battery lifetime which further affects the throughput QoS performance such as reduced packet loss, longer battery lifetime, reliability, and high data rates [1]-[35].

A review of the literature has shown that energy-efficient management algorithms have been proposed previously to address energy management in the 5G dense HetNets [1]-[35]. However, these existing algorithms do not satisfy the throughput QoS requirements such as high data rates, reduced latency, and reliability with regards to the three elements of energy consumption of the 5G radio access network (RANs). The 5G RANs govern the overall mobile communication networks for all the telecom operators in the world. These 5G RANs elements include specifically, the power consumed in the backhaul network (BHN), the power consumed in the SBSs, and that of the radio network controller (RNC) [1]-[35]. [1]-[35].

As a result, this paper proposed the Intelligent Throughput-based Sleep Control (ITSC) algorithm for balancing the trade-off between energy efficiency, throughput QoS requirements, and latency performance in the 5G dense HetNets. In the proposed ITSC algorithm, a deep neural network (DNN) is used to determine the cell capacity ratio for the SBSs. Hence, the SBSs cell capacity ratio was employed as decision criteria to put the SBSs into a sleep state. In addition, transferable payoff coalitional game theory was used in order to ensure real-time applications have a higher priority over non-real time applications. This was done in order to satisfy the throughput QoS requirements such as high data rates, reduced latency and packet loss, and reliability in the 5G dense HetNets.

The remainder of this paper is organised as follows: In Section 2, we provide the related works. In Section 3, we give the design of the proposed ITSC algorithm. In Section 4, we present the simulation results. In Section 5, we conclude the paper and provide future work.

## 2  Related Work

The review of literature has shown that many researchers have developed many algorithms for energy efficiency for the 5G dense HetNets. In this section, we discuss these existing algorithms and their limitations.

*Banerjee et al.* [13] proposed the use of the Energy Efficient Routing (EER) algorithm for Wireless Sensor Networks (WSNs) in order to prolong the battery lifetime of the sensor nodes. This technique defines the optimal size of clusters where data are aggregated properly, and data routing occurs in an energy-efficient way. In the EER algorithm, clusters are generated with the help of the L-system Cellular Automata (CA) scheme. The data routing path is also decided by the L-system CA scheme. The simulation result showed that the lifetime of the sensor nodes increased with respect to other existing energy conservation techniques. However, the EER algorithm failed to allocate higher bandwidth to real-time traffic which increased the number of packet loss and poor network throughput. The proposed ITSC algorithm reduced the number of packet losses by allocating higher bandwidth to real-time traffic using transferable payoff coalitional game theory.

*Karimpour and Abad* [14] proposed the New Routing (NR) algorithm which was designed by improving on the LEACH routing protocol. The NR algorithm was designed for energy saving, energy efficiency, and energy balancing in order to extend the service life of the WSNs. Serval simulation results showed that the NR algorithm extended the service life of the WSNs. However, the NR algorithm didn't improve the network throughput. This is because the NR algorithm failed to reduce the number of packet losses for real-time applications. The proposed ITSC algorithm improves the network throughput by minimalizing the number of packet losses. In the proposed algorithm, transferable payoff coalitional game theory was used in order to ensure real-time applications have a higher priority over non-real time applications. This was done in order to satisfy the throughput QoS requirements such as high data rates, reduced packet loss, and reliability in the 5G dense HetNets.

*Zhang et al.* [15] proposed the sleep mode activation scheme for small cells by the core network, which activates the sleeping SBS with the smallest large-scale channel fading to the active user. This scheme used the advantage of the spatial correlation in large-scale fading and formulates the channel estimation method as a linear mean square. The result obtained in their paper showed the effectiveness of the channel estimation method to estimate large-scale fading and saves transmit power over small cell networks.

In [16], *Soh et al.* modeled sleeping strategy and apply the tool from stochastic geometry to analyze the impact of load-aware sleeping strategy for the homogeneous macro cell and heterogeneous networks. Based on the simulation results their model improved the lifetime of the homogeneous macro cell and heterogeneous networks.

*Tang et al.* [22] proposed an energy efficient and reliable routing algorithm based on DSevide cetheory (DS-EERA). Firstly, their DS-EERA established three attribute indexes as the evidence under considering the neighboring nodes' residual energy, traffic, the closeness of its path to the shortest path, etc. Then they adopted the entropy weight method to objectively determine the weight of three indexes. After establishing the basic probability assignment (BPA) function, the fusion rule of DS evidence theory was applied to fuse the BPA function of each index value to select the next hop. Finally, each node in the network transmits data through this routing strategy. Simulation results showed that DS-EERA can effectively prolong the network lifetime.

However, the schemes proposed in [15], [16], [22] could not improve the network throughput since their schemes experienced a high number of packet losses. This is because their schemes failed to allocate higher bandwidth to real-time traffic. The proposed ITSC algorithm used transferable payoff coalitional game theory to diminish packet loss in order to improve network throughput.

The review of the literature indicated that existing energy efficiency algorithms do not satisfy the throughput QoS requirements such as lessened latency and packet loss, reliability, and high data rates. In the next section, we present the design of the proposed ITSC algorithm in order to solve the identified problem.

## 3 Design of Intelligent Throughput-based Sleep Control (ITSC) Algorithm

The Intelligent Throughput-based Sleep Control (ITSC) algorithm was designed by using a Deep Neural Network (DNN) to determine the cell capacity ratio for the SBSs. Hence, the SBSs cell capacity ratio was employed as decision criteria to put the SBSs into a sleep state. In addition, transferable payoff coalitional game theory was used in order to ensure real-time applications have a higher priority over non-real time applications. This was done in order to satisfy the throughput QoS requirements such as high data rates, reduced latency, and packet loss, and reliability in the 5G dense HetNets (see Fig. 1).

### 3.2.1 Deep Neural Network

DNN was used to calculate the cell capacity ratio for each SBSs. Different traffic types namely, real-time applications (i.e., voice traffic and video streaming) and non-real time applications (i.e., HTTP and e-mail) were used as input parameters to calculate the cell capacity ratio for each SBSs. We used a packet size of 84480- and 78022-bytes' online video conferencing and voice call for real-time applications. We further used an FTP packet size of 500 bytes for email services and 1000 bytes for HTTP services in the 5G dense HetNets (see Fig. 2).

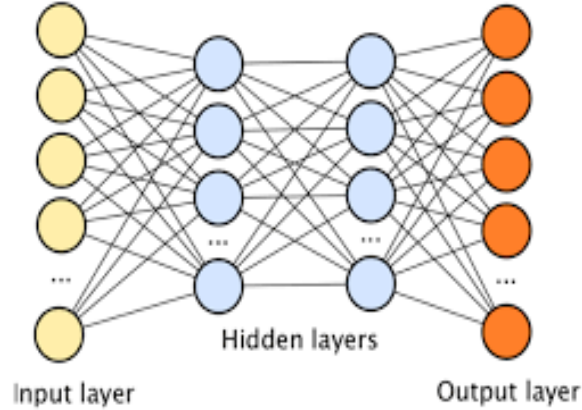

**Fig. 2.** Deep neural network architecture [21]

The cell capacity ratio for each SBSs was calculated using equation (1).

$$CC_i = P_1 + P_2 + P_3 + P_n \quad (1)$$

Where $CC_i$ is the cell capacity of each SBSs, and $P_1$ is any traffic type. After calculating the cell capacity of each SBSs and if the $CC_i = 0$ of any small base station (SBS), that SBS will go in a sleep state in order to save energy for each SBS. If the $CC_i = 0$, it means that there is no mobile node (MN) or user equipment (UE) that is connected to the SBS (see equation (2)).

$$SBS_j = 0 \quad (2)$$

While if $CC_i \neq 0$, it means that there are UE connected to that SBS (see equation (3)).

$$SBS_j = UE_1 + UE_2 + UE_3 + UE_n \quad (3)$$

Therefore, the energy efficiency ($EE_i$) is defined by the total cell capacity of each SBSs per total power consumption ($PC_{total}$) (see equation (4)).

$$EE_{SBS_j} = \frac{CC_1 + CC_2 + CC_3 + CC_n}{PC_{total}} \quad (4)$$

### *3.2.2 Transferable Payoff Coalitional Game Theory*

Transferable payoff coalitional game theory was used in this paper in order to give real-time application higher priority over non-real time traffic. This was done in order to achieve high data rates, reduced latency, and reliability for real-time applications in the 5G HetNets. For real-time traffic, we used a packet size of 84480 and 78022 bytes for online video conferencing and voice calls (see equation (5)). For non-real-time traffic, we further used an FTP packet size of 500 bytes for email services and 1000 bytes for HTTP services (see equation (5)).

$$\begin{aligned} &\text{If } P_n \geq 78022 \\ &\text{BW}_H = \left[ BW_{total} - \sum_i^N G_i^H \right] \\ &\text{elseif } P_n \leq 500 \\ &BW_L = \left[ \frac{BW_{total} - \text{BW}_H}{\sum_i^N R_i^M} \right] \end{aligned} \quad (5)$$

### *3.2.3 Intelligent Throughput-based Sleep Control (ITSC) Algorithm*

The proposed ITSC algorithm was designed by integrating DNN and transferable payoff coalitional game theory in order to satisfy the throughput QoS requirements such as high data rates, reduced latency, and reliability in the 5G dense HetNets (see Algorithm 1).

**Algorithm 1: Intelligent Throughput-based Sleep Control Algorithm**

**Input**
1. $P_1, P_2, P_3, P_n$ // Traffic types
2. $UE_1, UE_2, UE_3, UE_n$ // user equipment

**Process**

3. while $P_i \neq 0$
4. $CC_i = P_1 + P_2 + P_3 + ...P_n$
5. while $CC_i \neq 0$
6. $SBS_j = UE_1 + UE_2 + UE_3 + ...UE_n$
7. $EE_{SBS_j} = \dfrac{CC_1 + CC_2 + CC_3 + ...CC_n}{PC_{total}}$
8. if $EE_{SBS_j} = 0$
9. $SBS_j$ //will go to sleep state
10. else
11. $SBS_j$ // will remain active

$\quad$ If $P_n \geq 78022$

$$BW_H = \left[ BW_{total} - \sum_{i}^{N} G_i^H \right]$$

$\quad$ elseif $P_n \leq 500$

$$BW_L = \left[ \dfrac{BW_{total} - BW_H}{\sum_{i}^{N} R_i^M} \right]$$

12.
13. $\quad$ end elseif
14. $\quad$ end if

**Output**

15. $\quad\quad\quad\quad$ return $BW_H / BW_L$
16. $\quad$ end if
17. end while
18. end while

This section presented the design of the proposed ITSC algorithm in order to improve the network throughput and satisfy the throughput QoS requirements such as diminished packet loss, reliability, and high data rates. In the next section, we present the evaluation of the proposed ITSC algorithm.

## 4 Simulation Results

In this paper, we implemented the proposed ITSC algorithm using the IEEE 802.11ac model developed using NS-2 version 35. We ran NS-2 on the Linux Ubuntu 14.04 operating system, 512 RAM, and 10 gigabytes of storage. In addition, we installed an LTE patch in order to set up LTE nodes and the LTE base stations. We further installed the Ns2viop patch to allow nodes to send and receive voice/video packets. We used the Tool Command Language (Tcl) script to simulate the network topology as given in Appendix B. In addition, since NS-2 uses C++ to implement network algorithms, it was easier to implement DNN and transferable payoff coalitional game theory methods as a result. We configured a network topology of 500mX500m with five fixed located SBSs (labelled SBS1, SBS2, SBS3, SBS4, and SBS5), and 10 randomly located nodes (see Fig. 3). We configured the simulation to start transmitting packets at 10 seconds and stop transmitting at 100 seconds.

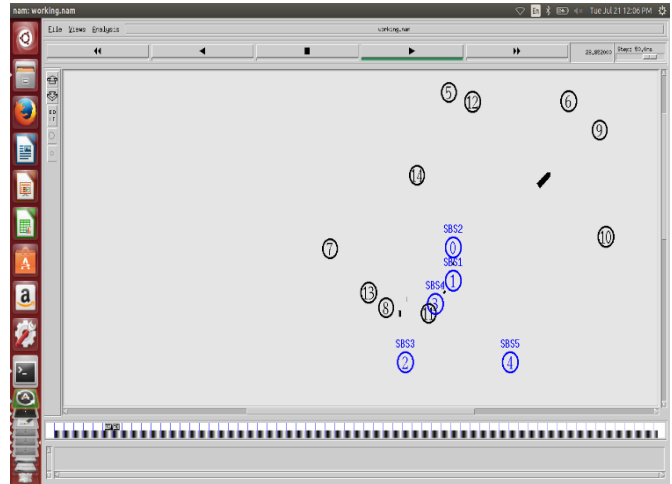

**Fig. 3.** NAM simulation scenario

In addition, we used the Ad-hoc On-Demand Distance Vector (AODV) routing protocol; this is because the AODV routing protocol enables dynamic, self-starting routing between MNs that are communicating. AODV creates routes only when they are needed, which makes it easier to be implemented in the 5G HetNets. In addition, AODV supports both unicast and multicast routing packet transmissions. AODV builds routes between MNs when source nodes need to establish and maintain communication with other MNs on a wireless network. In addition, we used 802.11ac because it is a dual-band wireless technology and supports simultaneous connections on both the 2.4 GHz and 5 GHz Wi-Fi bands. Furthermore, 802.11ac offers backward compatibility to 802.11b/g/n, and bandwidth is rated up to 1300 Mbps on the 5 GHz band plus up to 450 Mbps on 2.4 GHz. We provided all the simulation parameters using Table 1.

**Table 1.** Simulation Parameters

| Parameters | Values |
|---|---|
| Routing Protocol | AODV |
| MAC Protocol | IEEE 802.11ac |
| Channel Type | WirelessChannel |
| Network Interface Type | WirelessPhy |
| Propagation Model | TwoRayGround |
| Antenna model | OmniAntenna |
| Queue Type | Queue/DropTail/PriQueue |
| Link Layer Type | LL |
| Number of nodes | 10 |
| Max packets in ifq | 100 |
| Mobility Mode | Random waypoint |
| Bandwidth | 50 Gpbs for each SBS |
| Simulation Area | 500mX500m |
| Simulation Time | 100 seconds |
| Agent Trace | ON |
| Router Trace | ON |
| Mac Trace | OFF |
| Movement Trace | ON |

For the radio propagation model, we chose to use a two-ray-ground reflection model, since it is more suitable for our computer simulation as it makes use of both the ground reflection and direct path for communication. Which for a small wireless network simulation, is more realistic than using a free-space model which adds simulation time. We configured our MNs to use an omnidirectional antenna for both transmission and reception. For the processing and forwarding of packets, we used the Link Layer (LL). We further used LossMonitor; a packet sink that checks for packet losses. In conclusion, the simulations were run for 100 seconds and were done ten times in order to obtain the most convincing results.

The proposed ITSC algorithm was compared with Energy Efficient Routing (EER) algorithm and the new routing (NR) algorithm. The EER algorithm uses a combination of media access control (MAC) and routing protocol for energy efficiency. The NR algorithm optimizes energy efficiency by reducing the number and total transmission distance in order to save energy.

*4.1.1 Average Packet Loss*

Packet loss occurs when one or more packets fail to reach their intended destination, during packet transmission. It is, therefore, important to minimize packet loss in order to ensure better QoS in the 5G dense HetNets.

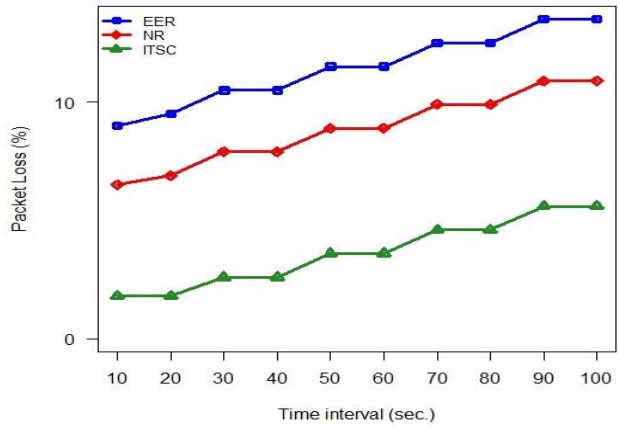

**Fig. 4.** Average packet loss

The proposed ITSC algorithm, the NR algorithm, and the EER algorithm have shown an average packet loss of 2.6%, 6.4%, and 7.7%, respectively (see Fig. 4). The proposed ITSC algorithm outperformed the NR algorithm and the EER algorithm. Mainly because the ITSC algorithm allocated higher bandwidth to the real-time applications using transferable payoff coalitional game theory. Hence, reduced packet loss while the NR algorithm and the EER algorithm failed to allocate higher bandwidth for real-time traffic experienced higher packet loss as a result.

*4.1.2 Average Energy Consumption*

Energy consumption is defined as the measurement of the amount of energy consumed during data transmission in a network. Hence, it is important to minimize the energy consumed in order to ensure better network throughput in the 5G dense HetNets.

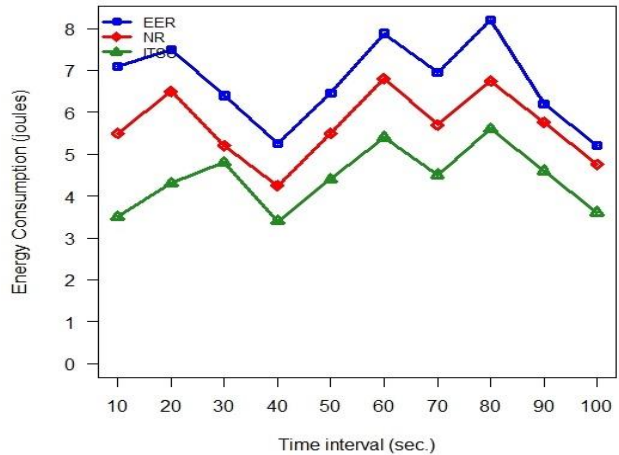

**Fig. 5.** Average energy consumption

We present the average energy consumption of the three algorithms compared as shown in Fig. 6. The simulation results illustrated that the proposed ITSC algorithm, the NR algorithm, and the EER algorithm experienced an average energy consumption of 3.4 joules, 4.9 joules, and 5.8 joules, respectively (see Fig. 5). The use of DNN in the proposed algorithm ensured that energy is saved on the SBS when there is UE connected to that SBS, hence, the ITSC algorithm provided a longer battery lifetime.
The proposed ITSC algorithm achieved that by putting the SBS into a sleep state when there is no UE currently connected to that SBS. The NR algorithm and the EER algorithm experienced a higher average energy consumption, mainly because both algorithms failed to put SBS into a sleep state when there was no UE connected to that SBS.

*4.1.3 Average Network Throughput*

Network throughput refers to the rate at which the data is transferred from a MN to its destination. We present the average network throughput of the three compared algorithms as shown in Fig. 6.

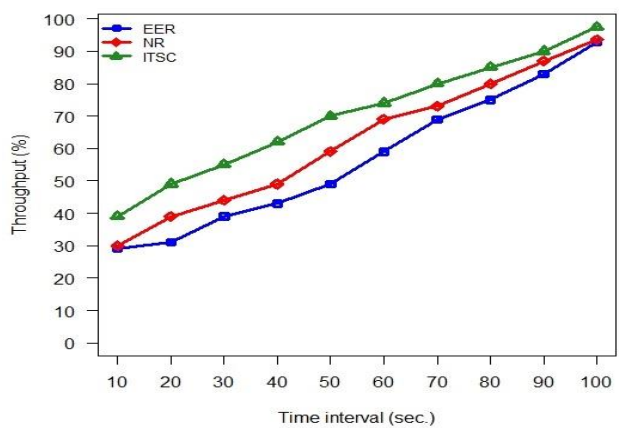

**Fig. 6.** Average network throughput

On average, the proposed ITSC algorithm, the NR algorithm, and the EER algorithm have shown 97.4%, 93.4%, and 92.3% throughput performance respectively, at a time interval of 100 seconds as exhibited in Fig. 6. The proposed ITSC algorithm produced higher network throughput mainly because the algorithm managed to reduce packet loss by using transferable payoff coalitional game theory. In addition, the use of DNN in the design of the proposed algorithm ensured that energy is saved on each SBS hence, the ITSC algorithm provided a longer battery lifetime. As a result, the proposed ITSC algorithm managed to satisfy the throughput QoS requirements such as reduced latency and packet loss, longer battery lifetime, reliability, and high data rates. The NR algorithm and the EER algorithm failed to allocate higher bandwidth to real-time applications, hence experienced higher packet loss which led to poor QoS.

## 5    Conclusion and Future Work

This paper presented the design of the proposed ITSC algorithm in order to satisfy the throughput QoS requirements such as reduced latency and packet loss, longer battery lifetime, reliability, and high data rates. The ITSC algorithm was designed by integrating DNN and transferable payoff coalitional game theory. DNN was used to determine the cell capacity ratio for the SBSs. Hence, the SBSs cell capacity ratio was employed as decision criteria to put the SBSs into a sleep state. In addition, transferable payoff coalitional game theory was used in order to ensure real-time applications have a higher priority over non-real time applications.

Several NS-2 computer simulation results illustrated that the proposed ITSC algorithm experienced an average packet loss of 2.6%, an energy consumption of 3.4 joules, and produced a network throughput of 97.4%. The proposed ITSC algorithm managed to satisfy the throughput QoS requirements such as packet loss, longer battery lifetime, reliability, and high data rates. In the future, the proposed algorithm will be integrated with security algorithms in order to ensure that data is protected while being transmitted from the source to the destination.

## Acknowledgments

The authors would like to thank the Tshwane University of Technology for financial support. The authors declare that there is no conflict of interest regarding the publication of this paper.